# Catalytic Resonance Theory: Kinetics and Frequency Response of Light-Promoted Catalysis

Paul J. Dauenhauer[1,2]*

[1] Center for Programmable Energy Catalysis (CPEC), University of Minnesota, 421 Washington Ave. SE, Minneapolis, MN, USA 55455.

[2] University of Minnesota, Department of Chemical Engineering & Materials Science, 421 Washington Ave. SE, Minneapolis, MN, USA 55455

* Corresponding author: hauer@umn.edu

**Abstract.** The illumination of catalytic surfaces with a continuous or pulsed stream of photons dynamically modulates surface chemistry for faster rates, higher conversion, or product selectivity control. To establish fundamental principles of dynamic photon-modulated catalysis, the photocatalytic conversion of a generic surface reaction was simulated to understand the kinetic implications of an independent stream of photons promoting surface product desorption. Simulations were conducted via the microkinetic model and kinetic Monte Carlo methods for the scenarios of a Poisson distribution, constant spacing between photons, and coordinated on/off pulsing of photon sources. The time-averaged photocatalytic rate at differential conditions for varying photon flux and temperatures indicated three kinetic regimes described by product thermal desorption control, surface reaction control, and an intermediate kinetic regime with a zero slope Arrhenius plot, consistent with a degree of rate control dominated by the photon arrival frequency (*i.e.*, per-site photon flux). The maximum photocatalytic rate occurred orders of magnitude above the Sabatier limit at the resonance frequency, identified as the photon arrival frequency that matched the surface reaction rate constant. Non-equilibrium photocatalytic conversion occurred when the rate of photon-promoted product desorption was comparable to the rate of thermal product desorption. Results indicated the general utility of photon-promoted catalysis, with negligible benefit for rate enhancement and quantum yield with photon source pulsing or chopping due to the inherent dynamic nature of photons from light sources.

**1.0 Introduction.** The maximum achievable catalytic rate, referred to as the Sabatier limit, has restricted the impact of catalysts in important reactions including water splitting, methanol synthesis, and ammonia synthesis.[1,2] An alternative approach to enhance chemical processing for faster catalysis, higher conversion, or improved selectivity to products[3] utilizes external dynamic stimuli to modulate the binding energy or reaction characteristics of surface species. The programmable condensation of charge,[4,5] the generation of local electric fields,[6,7] the modulation of electron density via strain,[8,9] or the imposition of photons have been shown to alter the characteristics of adsorbed molecules in energy-relevant reactions.[10,11] Periodic stimulation of the binding energy of reacting surface species accelerates the catalytic rate of turnover beyond the Sabatier limit, particularly when the stimulation frequency matches the kinetics of the rate-limiting surface processes (*i.e.*, the resonance frequency).[12,13] Energetic stimulation of the catalyst at this particular frequency maximizes the catalytic rate of turnover and efficiency with which the energy input associated with catalyst stimulation is used, defined as the 'turnover efficiency.'[14] Resulting from the mechanisms of energy input, catalyst stimulation can drive reactions to extreme extents of conversion far from equilibrium (*i.e.*, non-equilibrium conversion).[14,15]

While catalyst stimulation with charge, fields, and strain modify all or most surface adsorbates simultaneously, photons exhibit more precise perturbation of surface species through selective interaction with specific adsorbates and surfaces/particles. For example, methanol decomposition on Pt exposed to visible wavelength photons (440 nm) increased the catalytic turnover frequency by accelerating the desorption of carbon monoxide, reducing its time-averaged coverage under reaction conditions.[16] Periodic illumination lead to additional enhancements, improving the apparent quantum efficiency of photons. Recently, through isobaric measurements of carbon monoxide adsorption on Pt nanoparticles, Beck et al. demonstrated the non-thermal mechanism through which photons alter the presence of surface CO*.[17] Upon interacting with adsorbed carbon monoxide, non-thermal desorption facilitated by visible wavelength photons was described with a single rate expression distinct from thermal desorption kinetics that indicated a specific wavelength-adsorbate-catalyst interaction useful for controlling surface chemistry via programmable catalysis.

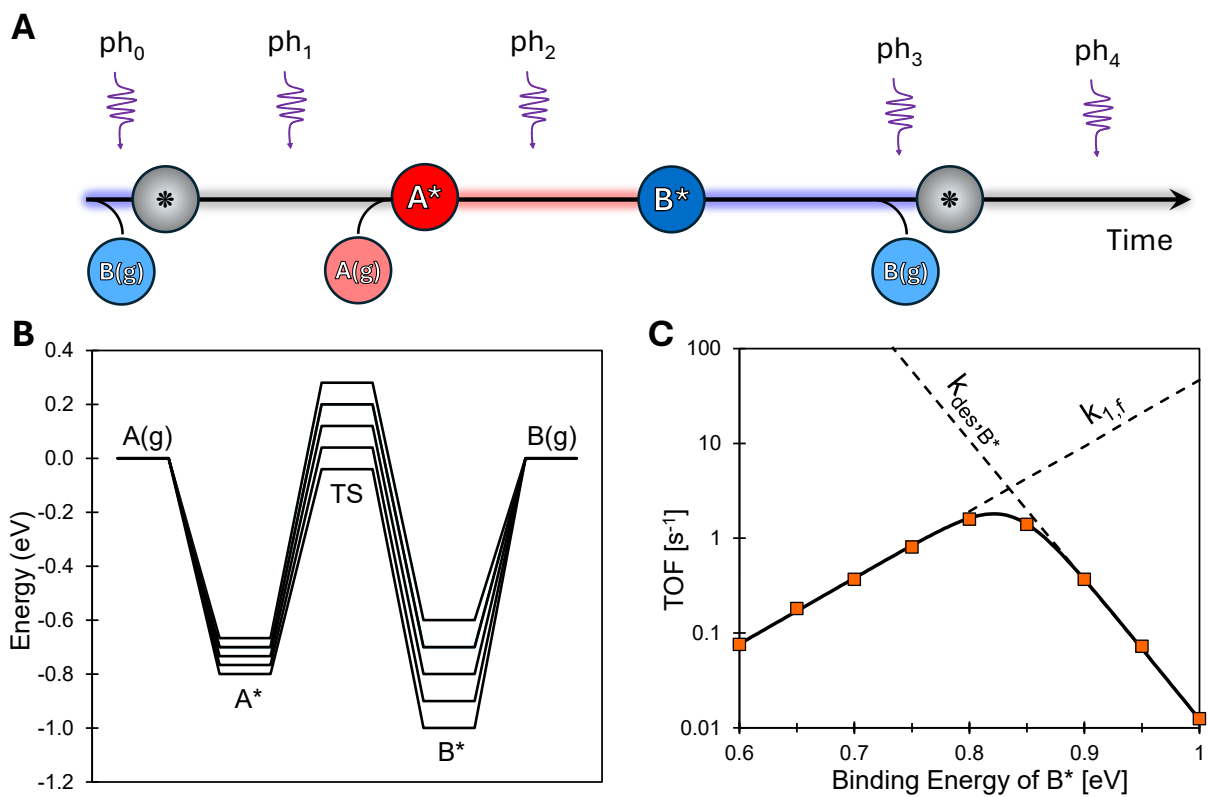

**Figure 1. Kinetics of Thermocatalysis for a Sabatier-Limited Reaction.** (A) The reaction of A(g) to B(g) through surface intermediates A* and B* occurs via stochastic elementary steps. Photons also arrive to the surface with an independent rate, sometimes arriving too early (ph$_1$ & ph$_2$), too late (ph$_4$) to promote photo-desorption of B* (ph$_3$). (B) The thermochemical reaction of A(g) to B(g) through A* and B* with variable binding energy determined by linear scaling: $\alpha = 0.7$, $\beta = 0.9$, $\gamma_{B/A} = 3$, and $\delta_{A-B} = 0.7$. (C) The turnover frequency (TOF) of the thermochemical catalytic reaction at 100 bar total pressure, 65 °C, and 1% conversion of A(g) to B(g) exhibits a volcano maximum at BE$_B$ of 0.82 eV of ~1.8 s$^{-1}$. Incident photons accelerate the catalytic rate to a maximum value, black, limited by the rate of surface reaction, $A * \rightarrow B *$. Simulated rates, orange, via kinetic Monte Carlo and black line via analytical kinetic solution.

Illumination of a catalytic surface is a dynamic perturbation of reacting surface species. Photon (ph) delivery via a continuous wave laser provides a flux of energy as high as ~2 watts/cm$^2$ of a selected wavelength that can be normalized to the number of active sites within the beam as an 'arrival frequency', $f_{arrival}$, as high as ~1,000 photons per site per second.[16,17]

$$f_{arrival} = \frac{Photon\ flux}{Active\ site\ areal\ density}\ [=]\ s^{-1} \quad (Eq.\ 1)$$

Photon-promoted desorption of adsorbates on metal catalyst surfaces is a fast event (femtoseconds to picoseconds) relative to the time between catalytic turnover events (seconds to kiloseconds),[17,18,19] indicating that photon-promoted desorption is a distinct perturbation event that can be considered as having an extreme duty cycle (10$^{-15}$ < DC < 10$^{-12}$ for $f_{arrival}$ ~1 Hz). These short-lived desorption events occur with a single photon promoting desorption of a single adsorbate by excitation of the electronic and vibrational states of CO* on Pt.[20] However, these photon-promoted desorption events occur only a small fraction of the time for molecules on the surface; the majority of the time adsorbates experience thermal mechanisms in the dark.

The independent kinetics of surface illumination and surface chemistry interact via the kinetics of photon-promoted surface reactions. A continuous wave laser generates a stream of photons separated temporally by a Poisson distribution.[21,22,23,24] From the perspective of the catalyst active site, the spacing between photons interacting with the surface is not predictable; some photons arrive soon after the preceding photon while others leave arrive much later. However, the average temporal spacing of photon arrival at the surface normalized to the number of sites is fixed as the arrival frequency, $f_{arrival}$. Simultaneously, catalytic reaction rates of multi-step surface reactions are quantified by the turnover frequency (TOF [=] s$^{-1}$]) defined as the average number of product moles formed per mole of active site per unit time.[25,26] The time between the occurrence of elementary steps in the catalytic cycle (*i.e.*, traversal of a transition state) also exhibits a Poisson distribution, with the elementary steps being defined by a rate coefficient (e.g., $k_1$ or $k_{des,B}$).[27,28,29] When combined in a photocatalytic process, surface illumination and surface reaction both exhibit stochastic events that are predictable and controllable only on average.

The illumination of catalyst surfaces provides a mechanism for external catalytic reaction stimulation with distinct temporal control. As depicted in **Figure 1A**, the A(g) to B(g) overall reaction proceeds exergonically downhill via the intermediates A* and B* with transitions between an overall average TOF determined by the rate determining step of A*→B* or B*→B(g). Simultaneously, photons approaching the surface arrive randomly in time after the initial photon ($ph_0$). If the selected wavelength of light only promotes the desorption of product B* and has no interaction with A*, then the photon arrival time relative to the surface reaction time determines the utility of the applied photon. Photons 1, 2, and 4 arrive to the surface either too early or too late to promote desorption of B*, while photon 3 arrives when the active site is occupied by B* and can potentially promote the formation of B(g). This indicates that the average flux of photons to the surface should be matched to the kinetics of the reaction, and the distribution of photons in time should be controlled as possible to minimize short and long timescales between photon arrival.

In this work, we leveraged this recent quantitative understanding[17] to evaluate the catalytic response of a unimolecular A-to-B surface reaction to incident photons. Photochemical events were simulated via the kinetic Monte Carlo (kMC) method and the method of microkinetic modeling for a range of general catalytic materials using linear scaling relationships between adsorbed reactants and transition states, such that photochemical response was evaluated for a class of catalytic materials characterized by the binding energy of B*, $BE_B$, as a materials descriptor. Interpretation of the resulting catalytic turnover frequencies for varying applied photon fluxes identified the kinetic parameters defining the frequency of incident photons leading to resonant photochemical behaviors, including catalytic rates beyond the Sabatier limit. Beyond this critical time scale, the quantum efficiency of photons driving non-thermal desorption monotonically decreased, resulting from a decreasing coverage of product species on the surface. Delivered at a frequency more than the intrinsic thermal rate of product desorption, incident photons enhanced the net rate of catalytic turnover such that non-equilibrium conversion was achieved. These insights are consistent with and understood through catalytic resonance theory.[14,15,30]

**Results and Discussion**. Reactions were simulated using a unimolecular A(g)-to-B(g) chemistry, proceeding on a catalytic surface through adsorbed intermediates A* and B* (**Figure 1B**). The relative adsorption energetics of A* and B* are described through linear free energy scaling parameters $\gamma_{B/A}$ and $\delta_{A-B}$, that define the relative change in the adsorption energetics of B* to A* and the equivalent binding energy, respectively. Transition state energies are dictated by $\alpha$ and $\beta$, which represent the change in the activation energy with surface reaction energy, and the thermal activation energy, respectively. Scaling relation values were selected in prior works with intermediate ($\gamma_{B/A}$ = 3, and $\delta_{A-B}$ = 0.7) and transition state ($\alpha$ = 0.7, $\beta$ = 0.9 eV) linear scaling parameters with moderate characteristics representing a broad range of reactions.[13,30] An overall standard state reaction free energy of zero ($\Delta G_{rxn}^0$ = 0) was selected, dictating an equimolar reaction environment at thermal equilibrium ($y_A = y_B$). Sabatier volcanos have been shown to exhibit robust prediction of catalytic performance for minimal variance in the general scaling parameters, consistent with the identified variability from DFT.[31] With intermediate binding scaling of $\gamma_{B/A}$ = 3 selected, the binding energy of B* for different catalytic materials varied more than A*, identified here with the descriptor of the binding energy of $BE_B$. For materials with $BE_B > \delta_{A-B}$ (0.7 eV), B* binds more strongly than A*, resulting in a higher surface coverage of B* ($\theta_B > \theta_A$).

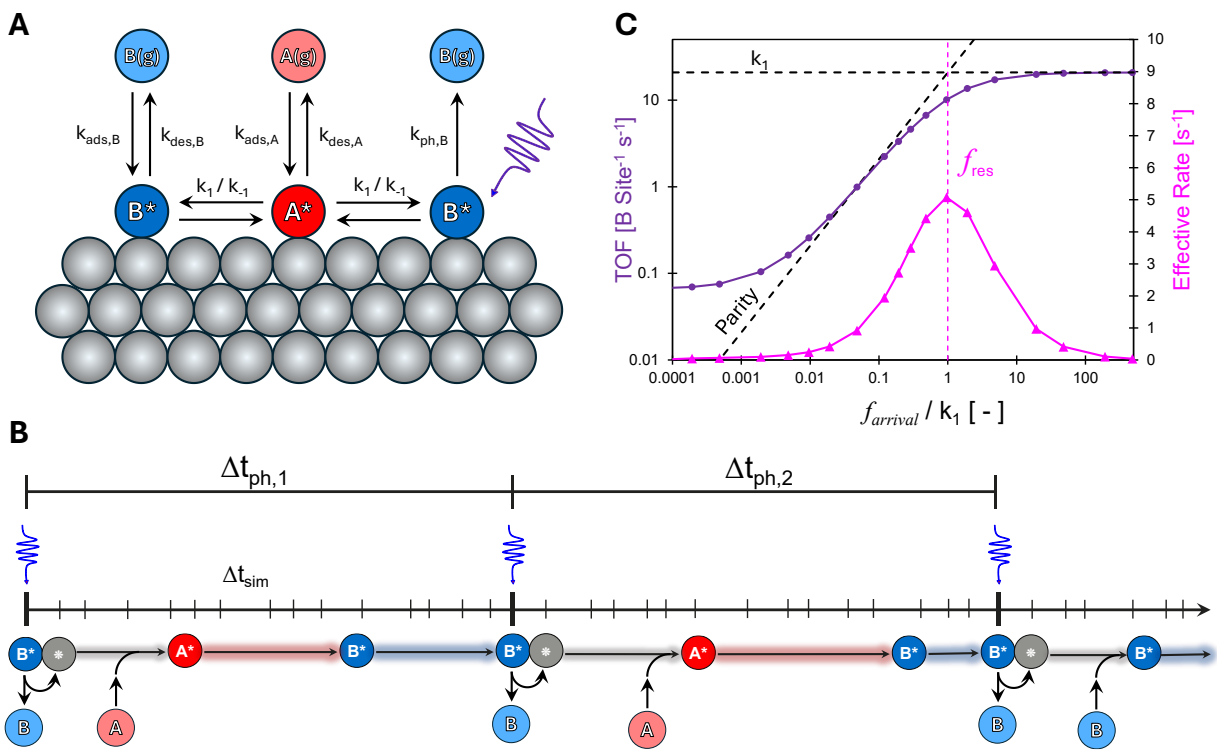

**Figure 2. Kinetics of Programmable Photocatalysis with Variable Photon Arrival Frequency. (A)** The reaction of A(g) to B(g) through A* and B* with thermocatalytic and photocatalytic rates. **(B)** The simulated kinetic Monte Carlo method exhibits variable time steps with the catalyst switching between states of ∗, A*, and B*. **(C)** Catalytic turnover frequency, $TOF_B$, for varying photon arrival frequency relative to parity (dashed line) at $BE_B$=0.95 eV, 100 atm, 338 K, and 1% conversion of A(g). The effective rate, pink, exhibits a maximum at the catalytic resonance frequency, $f_{res}$ for $BE_B$ = 0.95 eV.

The thermocatalytic turnover frequency to form B(g), $TOF_B$, was calculated via the analytical solution of Eq. 2 at differential conditions of $X_A$ of 1%, 65 °C, and total pressure of 100 bar with equilibrium and kinetic coefficients defined by the scaling relationships,

$$TOF_B = \frac{k_1 K_A P_A}{1 + K_A P_A + k_1 K_A P_A / k_{des,B}} \quad \text{(Eq. 2)}$$

In addition, the thermocatalytic rates of materials were simulated via kinetic Monte Carlo (kMC), and calculated $TOF_B$ at varying binding energies of B* (**Figure 1C**) agreed with the analytical solutions. Kinetic Monte Carlo simulations adopted a more probabilistic approach in contrast to a mean-field microkinetic model. Full simulation details are provided in the methods section and supplemental information.

The selected chemical scaling parameters yield a Sabatier volcano (**Figure 1C**) with a distinct maximum in catalytic turnover frequency of ~1.8

B(g)·s⁻¹·site⁻¹ at $BE_B$ of 0.82 eV. For materials with $BE_B$ > 0.82 eV, the overall thermocatalytic rate of turnover was controlled by the rate of desorption of B* ($k_{des,B}$); alternatively, the overall rate limitation for materials with $BE_B$ < 0.82 eV was the forward surface reaction, $k_1$, to form B* from A*. The rate limitations of the Sabatier volcano were defined by the transition in rate-controlling elementary steps identified by the dashed lines of **Figure 1C**. The Sabatier peak is therefore characterized by a transition in the degree of rate control between these two elementary steps.

While the thermocatalytic reaction is limited by the Sabatier volcano, external stimulation via photons provided an additional pathway to accelerate the reaction. Desorption of B* upon interaction with a photon was assumed (**Figure 2A**), which yielded an additional unidirectional reaction path to accelerate the removal of B* from the surface defined by rate constant $k_{ph,B}$,

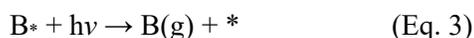
$$B_* + h\nu \rightarrow B(g) + * \quad \text{(Eq. 3)}$$

This addition of an irreversible photodesorption elementary step was inspired by Beck et al., where such an approach accounted for the time-averaged coverage of carbon monoxide on supported Pt nanoparticles.[17] The combination of thermocatalytic and photodesorption steps was simulated to understand the influence of photons at an active site on the overall rate of the catalytic cycle, under a desorption limited regime ($BE_B$ = 0.95 eV) (**Figure 2B**). The rate constant of a photochemical event, such as photodesorption, is typically defined as a product of quantum efficiency and the rate at which photons are delivered. To define the rate of photons, the photon flux was divided by the area density of active sites to yield a photon arrival frequency, $f_{arrival}$, defining the number of photons impacting the site per unit time (Eq. 1). In effect, this normalization of photon flux per site as opposed to per surface area represents the average rate at which a catalytic site experiences a photon.

The combined thermocatalytic and photodesorption events driving catalytic turnover ($TOF_B$) were therefore calculated as a function of the photon arrival frequency, where a sigmoidal dependence on the photon arrival frequency was observed (**Figure 2C**). Across the span of photon arrival frequency, the simulated $TOF_B$ exhibited three kinetic regimes. At low applied photon arrival frequencies ($f_{inc}/k_1$ < ~0.005), the overall catalytic rate of turnover, was relatively unchanged and remained defined by the Sabatier volcano ($TOF_B$ ~ 0.07 B(g)·s⁻¹·site⁻¹). At high frequencies ($f_{inc}/k_1$ > 1), the catalytic rate approached the maximal rate of surface reaction dictated by the forward kinetic rate constant ($k_1$ = 20.9 B(g)·s⁻¹·site⁻¹). In the intermediate kinetic regime, the simulated $TOF_B$ exhibited parity (dashed line) with the surface reaction normalized photon arrival frequency ($f_{inc}/k_1$); the overall catalytic rate of turnover to B was linearly proportional to the rate at which photons arrived at the active site.

The transition to the high photon arrival frequency kinetic regime ($f_{inc}/k_1$ > 1) is identifiable by a maximum in the effective catalytic rate. The effective rate has been previously defined as a product of the catalytic turnover frequency, $TOF_B$, and the turnover efficiency.[13,14]

$$TOF_{eff} = TOF_B \left(\frac{TOF_B - TOF_{SS}}{f_{inc}}\right) \quad \text{(Eq. 4)}$$

The effective rate exhibits a maximum at the resonance frequency by definition,[13] which in this case equals the forward surface reaction rate coefficient, $k_1$. The effective catalytic rate, $TOF_{eff}$, increases with increasing $TOF_B$ up to the resonance frequency but then decreases above $f_{res}$ as photons arriving at the surface no longer enhance the rate (i.e., they arrive to empty sites or sites with A*). For materials with stronger binding energies (i.e., higher $BE_B$), the forward surface reaction rate constant, $k_1$, is larger per scaling relations, shifting the resonance frequency to higher photon arrival frequencies, $f_{arrival}$, (Figure S5).

The rate enhancement resulting from photon flux to the catalyst surface can be interpreted in context of the Sabatier volcano (**Figure 3A**). Simulated catalytic rates for high photon arrival frequencies ($f_{arrival}/k_1$ >> 1) depicted as black squares, ■, all exist on the dashed line extending from the low binding energy side of the Sabatier volcano, which is defined by the forward rate constant of the surface reaction ($k_1$). The catalytic rates resulting from varying photon arrival frequencies are thus bound by the purely thermocatalytic rate of the Sabatier volcano, $TOF_{Thermal}$, and the forward surface reaction rate, $k_1$.

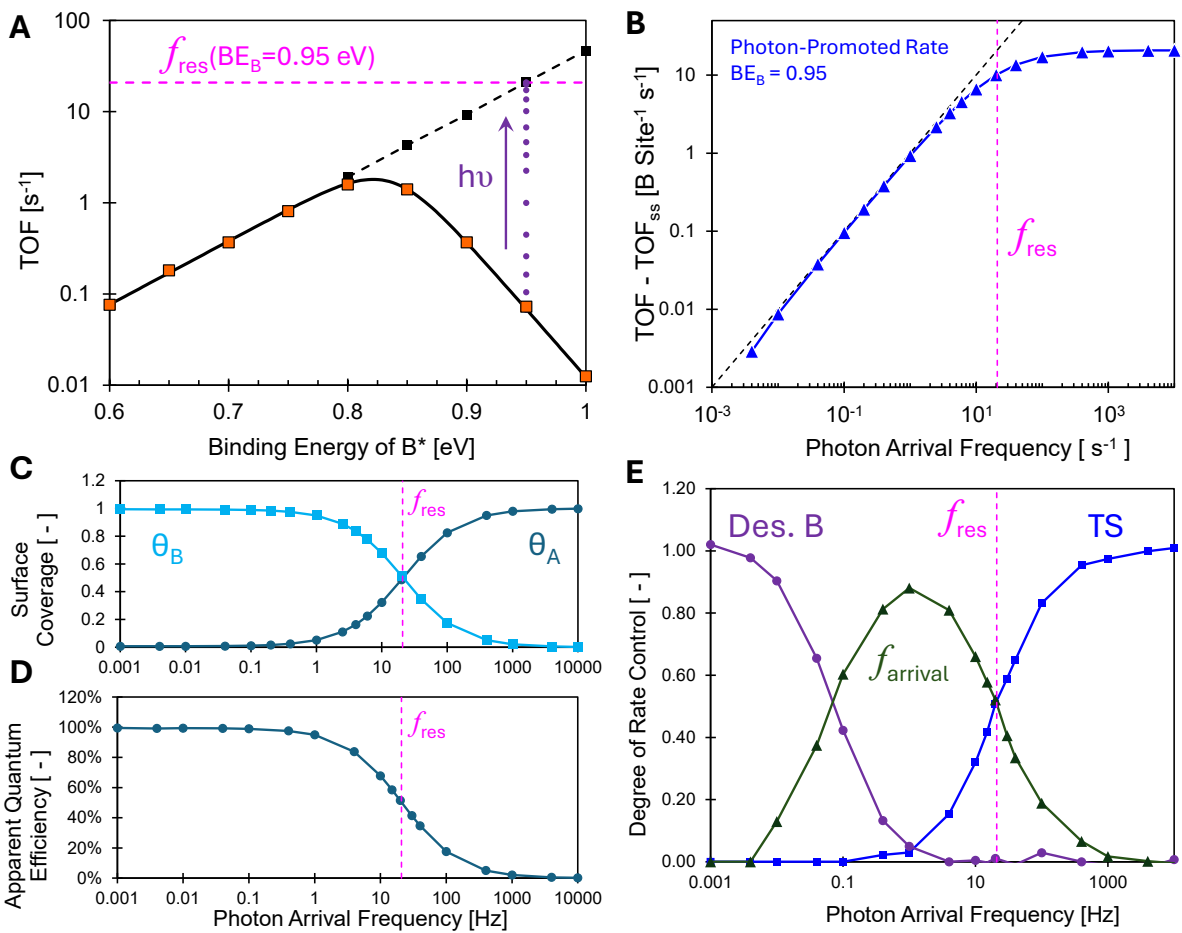

**Figure 3. Kinetics of Programmable Photocatalysis with Variable Photon Arrival Frequency. (A)** The reaction of A(g) to B(g) through A* and B* with variable binding energy determined by linear scaling: $\alpha = 0.7$, $\beta = 0.9$, $\gamma_{B/A} = 3$, and $\delta_{A-B} = 0.7$ at 100 atm, 338 K, and 1% conversion of A(g). The KMC-simulated turnover frequency (TOF) of the catalytic reaction, orange ■, exhibits a volcano maximum of ~1.8 B(g)·s$^{-1}$·site$^{-1}$. Applied photons accelerate the catalytic rate to a maximum value, black ■, limited by the rate of surface reaction, $A* \rightarrow B*$. **(B)** The photon-promoted rate (TOF-TOF$_{ss}$, blue) for BE$_B$ = 0.95 exhibits parity with the photon arrival frequency up to the resonance frequency. **(C,D,E)** The surface coverage, apparent quantum efficiency, and degree of rate control of the overall photocatalytic reaction for varying photon arrival frequencies, $f_{arrival}$, at BE$_B$=0.95 eV, 100 atm, 338 K, and 1% conversion of A(g).

For a catalytic material defined by BE$_B$ of 0.95 eV (purple circles ●, **Fig. 3A**), the thermocatalytic rate of turnover (1.8 B(g)·s$^{-1}$·site$^{-1}$) was limited by the desorption of B* to form B(g). Deviation from the thermocatalytic rate arises from the enhanced rate of desorption of B* to B(g) promoted by photons, approaching a maximum of 20.87 B(g)·s$^{-1}$·site$^{-1}$ under simulated illumination. The contribution of photons to the overall rate of turnover was calculated by subtracting the time-averaged thermocatalytic rate in the absence of photons (1.8 B(g)·s$^{-1}$·site$^{-1}$ @ 0.95 eV, **Figure 3B**), where the photon-driven rate of turnover (TOF-TOF$_{Thermal}$) was proportional to the photon arrival frequency up to the resonance frequency of 20.87 Hz. In these intermediate arrival frequencies, the photon-driven catalytic rate of turnover exhibited a first order dependence on photon flux per site. Direct overlap with the parity line indicates that photons are used efficiently.

At arrival frequencies above the resonance frequency, a departure from the efficient use of photons was observed (**Figure 3B**). Photons arriving at the active site no longer interacted with B* but rather other catalyst states such as A* or a vacant site, *, resulting in no benefit to the rate of catalytic turnover. This loss of efficiency was apparent when considering calculated surface coverages (**Figure 3C**); a distinct transition between high surface coverage of B* to high surface coverage of A* was observed at the resonance frequency. Only a negligible number of active sites are vacant, ∗, under the simulated conditions, primarily due to the high operating pressure of 100 atm. The reduced utility of photons at these higher arrival frequencies was quantified as the apparent quantum efficiency (**Figure 3D**). At low arrival frequencies, photons arriving at the active site primarily encounter B* and result in a catalytic turnover. At high photon arrival frequencies, photons at the surface primarily encounter A* and have no catalytic benefit, thereby reducing the quantum efficiency.

The transition between the three kinetic regimes and quantum efficiency was also rationalized through the catalytic degree of rate control (DRC).[30,32,33] For each rate constant $k_i$, the sensitivity, $S_i$, was calculated as,

$$S_i = \left(\frac{\partial \ln\langle TOF_B \rangle}{\partial \ln k_i}\right)_{k_{i \neq j}} \quad \text{(Eq. 5)}^{[30]}$$

The DRC was then calculated as the sum of the sensitivities for the forward, $S_i$, and reverse, $S_{-i}$, rate constants. The degree of rate control attributed to the photon arrival frequency was similarly calculated as the derivative of the log of the time-averaged catalytic rate relative to the log of the photon arrival frequency,

$$DRC_f = \frac{d \ln\langle TOF_B \rangle}{d \ln f_{arrival}} \quad \text{(Eq. 6)}^{[30]}$$

By this definition, the $DRC_f$ quantifies the influence of *temporal characteristics* of photons arriving at the active site on the rate of formation of B(g); this is determined by the parameter, $f_{arrival}$, set as an input to the experiment or simulation.

As shown in **Figure 3E**, the three kinetic regimes are associated with three regimes of degree of rate control. For low photon arrival frequencies, the rate of reaction is predominately controlled by the thermocatalytic desorption of B* to B(g), consistent with the Sabatier volcano. Photons arrive too infrequently relative to thermocatalytic desorption, which is sufficiently fast to define the overall rate of catalytic turnover to B(g). Alternatively, at arrival frequencies above the resonance frequency, the overall catalytic rate is defined overwhelmingly by the surface reaction transition state energy. Photons arrive at active sites frequently (relative to reaction timescales) such that the conversion of A* to B* cannot keep up; when B* eventually occupies a site, it experiences a photon almost instantly and desorbs. The intermediate kinetic regime is the transition between extreme kinetics; photodesorption proceeds faster than thermal desorption, but photons are not yet arriving at active sites more frequently than A* can convert to B*. Thus, the catalytic reaction is predominately controlled by the photon arrival frequency.

The rate of catalytic turnover assisted by photons is frequently reported to exhibit strong temperature dependence, despite the non-thermal nature of photochemical events.[17] We therefore sought to understand the consequence of varying temperatures, considering the influence of photons on the apparent activation (**Figure 4A**). Again, the simulated total catalytic rate, $TOF_B$, at 100 bar and $X_A$ of 1% exhibited three kinetic regimes with varying temperature at the applied arrival frequencies of 1 Hz (green), 10 Hz (blue), and 100 Hz (purple). At elevated temperatures (low inverse temperature), all simulated arrival frequencies exhibited the same $TOF_B$ defined by the thermocatalytic rate controlled by the thermal rate of B* desorption, $k_{desB}$. The apparent activation energy under elevated temperature was therefore equal to the barrier to desorption (0.95 eV). Alternatively, at low temperatures (high inverse temperatures), all simulated arrival frequencies exhibited the same $TOF_B$ defined by the forward reaction rate constant, $k_1$. The surface reaction rate limitation is apparent through comparison with the photon-promoted catalytic rate, which exhibits identical low temperature $TOF_B$ kinetic behavior (**Figure 4B**). As the temperature increases, the catalytic rate converting A* to B* eventually catches up to the applied photon arrival frequency, resulting in a transition from high surface coverage of A* at low temperatures to high surface coverages of B* at high temperatures (**Figure 4C**).

The unique kinetic behavior in the Arrhenius plots of **Figure 4** exists within the intermediate kinetic regime of moderate inverse temperatures where all three applied photon arrival frequencies exhibit nearly zero apparent barrier. For these conditions, the rate is entirely controlled by the applied photon arrival frequency, resulting in the $TOF_B$ matching the horizontal black-dashed lines corresponding to 1, 10, and 100 Hz. This region of zero apparent barrier is the same intermediate kinetic regime identified under isothermal conditions, with a relatively large photon arrival frequency degree of rate control ($DRC_f$, **Figure 3**). The zero apparent barrier therefore results from the non-thermal nature of selecting the photon arrival frequency, $f_{arrival}$, as previously observed for other types of perturbations.[30]

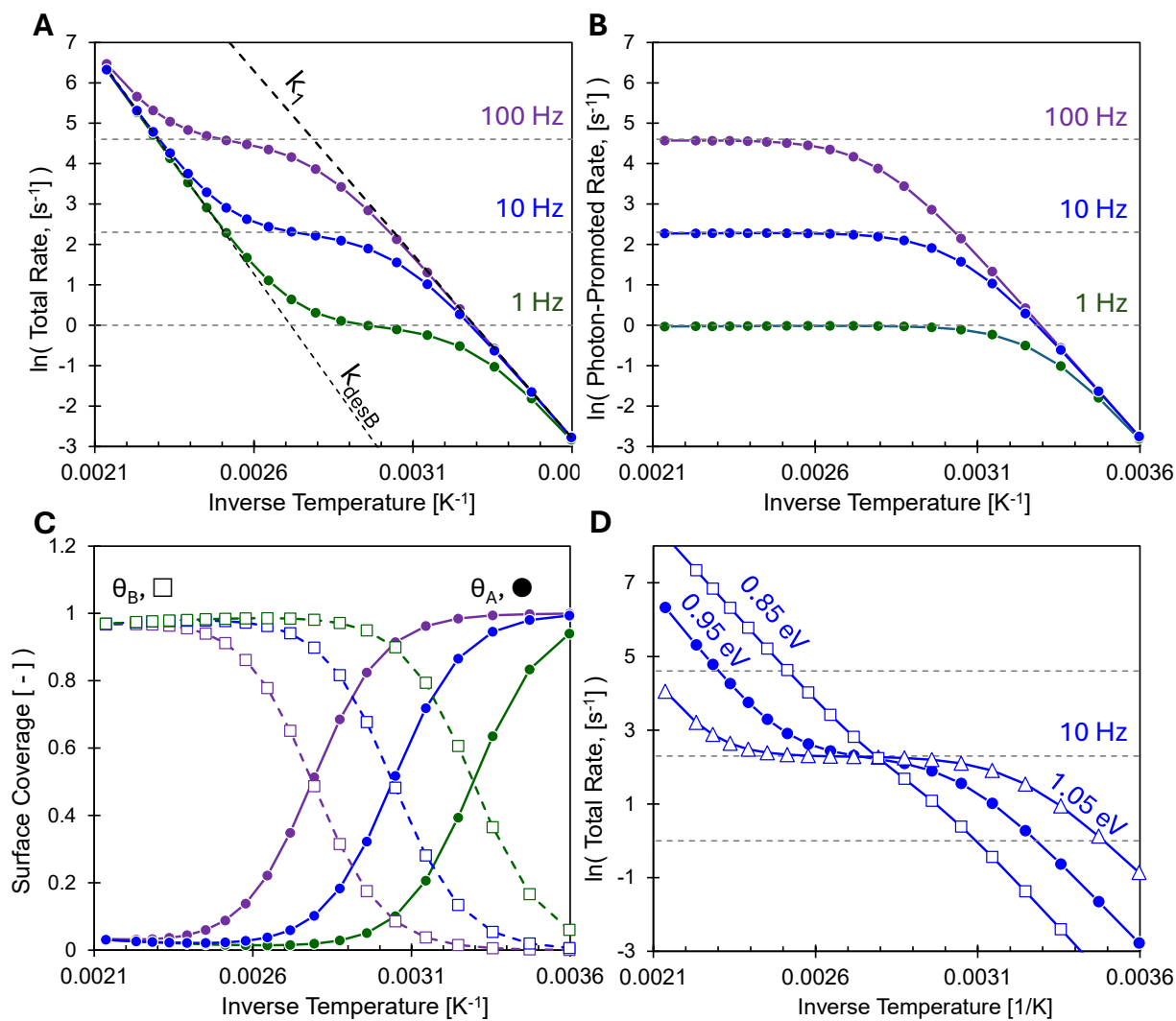

**Figure 4. Arrhenius Kinetics of Photon-Promoted Catalysis. (A)** Arrhenius plot of the total formation rate of B(g) for $BE_B$ of 0.95 eV at photon arrival frequencies of 1 Hz (green), 10 Hz (blue), and 100 Hz (purple) for temperatures from 278 to 476 K. **(B)** Arrhenius plot of the photon-promoted rate of formation of B(g) at 1 Hz, green, 10 Hz, blue, and 100 Hz, purple. **(C)** The surface coverage of A*, square, and B*, circle, for varying inverse temperature. **(D)** The Arrhenius plot of total rate versus inverse temperature for an applied photon arrival frequency of 10 Hz for materials with $BE_B$ of 0.85, 0.95, and 1.05 eV.

*Photocatalytic Rate Orders.* The time-averaged turnover frequency was simulated for varying photon arrival frequency as a function of varying partial pressure of A(g) between $10^{-8}$ and 1,000 bar while maintaining the pressure of B(g) at $10^{-10}$ bar and a temperature of 348 K for a material characterized with $BE_B$ of 0.95 eV. As shown in **Figure 5A**, the TOF increased proportionally for all considered photon arrival frequencies up to about $10^{-6}$ to $10^{-4}$ bar. Above this pressure range, the TOF was the same for each considered photon arrival frequency. The transition in TOF was associated with a shift in the reaction rate order of A(g). As depicted in **Figure 5B**, all considered simulations over the full range of photon arrival frequencies shift from first order to zero order in A(g) as the pressure of A(g) increased.

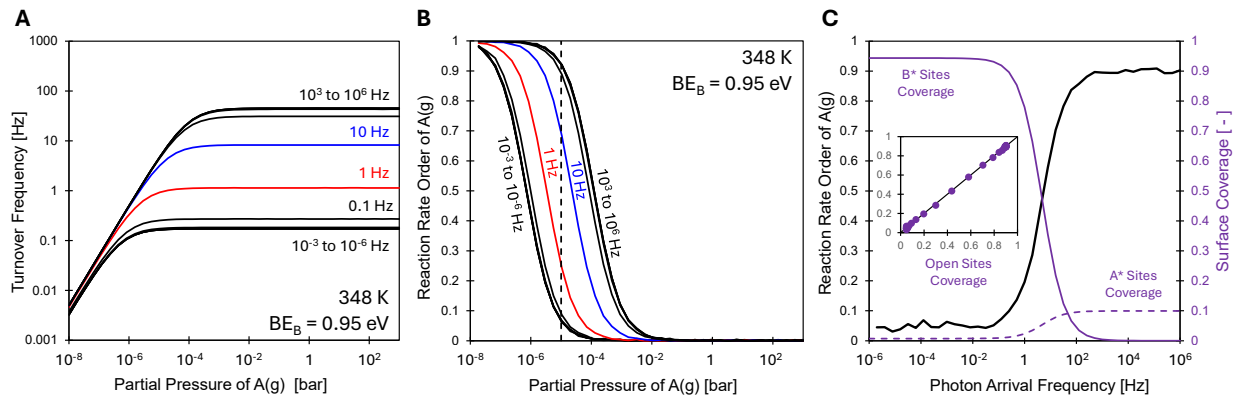

**Figure 5. Reaction Rate Orders of Photon-Promoted Catalysis.** The time-averaged turnover frequency **(A)** of the formation of B(g) and the reaction rate order of A(g) **(B)** for varying partial pressure of A(g) and photon arrival frequency at 348 K and a BEB of 0.95 eV. The pressure of B(g) was maintained at $10^{-10}$ bar for all simulations. **(C)** The reaction rate order of A(g) was calculated for a fixed pressure of $10^{-5}$ bar with varying photon arrival frequency at fixed temperature of 348 K and $BE_B$ of 0.95 eV. For all simulations, the pressure of B(g) was maintained at $10^{-10}$ bar. The surface coverages of B* and A* and open sites (*) were calculated for all applied photon frequencies; parity was observed (inset) between the reaction rate order of A* and the open site surface coverage.

To better understand the relationship between reaction rate orders and the surface conditions, a single pressure of A(g) from **Figure 5B** of $10^{-5}$ bar (vertical dashed line) was plotted in **Figure 5C** for a range of photon arrival frequencies ($10^{-6}$ to $10^6$ Hz). As shown, the reaction rate order of A(g) is ~0.05 up to ~0.1 Hz, then exhibits a transition up to ~0.9 for 100 Hz and higher frequencies. To understand this transition, the surface coverages of A*, B*, and open sites * were calculated and plotted. The surface coverage of all species change for the same transition frequencies as the reaction rate order of A(g). In particular, the reaction rate order of A(g) exhibits parity with the surface coverage of open sites on the catalyst surface, indicating the clear connection between the pressure of A(g) and the number of open surface sites on the overall rate of reaction.

*Non-Equilibrium Photocatalysis.* The photochemical system was also simulated at variable conversion to determine the time-averaged conversion at long reaction times. The time-averaged conversion was identified as the simulated conversion yielding a net zero total rate, $TOF_B$, such that all thermochemical and photocatalytic steps resulted in equal forward and reverse overall rates (**Figure 6**). The conversions were calculated for a range of materials with different $BE_B$ existing on both sides of the Sabatier volcano (0.75 eV < $BE_B$ ≤ 0.95 eV) for varying applied photon arrival frequencies (**Figure 6A**). For low photon arrival frequencies, all simulations exhibited a time-averaged conversion matching that of thermochemical equilibrium ($X_A$ = 50%). At high arrival frequencies ($f_{arrival}$ > 10,000 Hz), all simulated conditions exhibited non-equilibrium conversions approaching near quantitative conversion ($X_A$ > 99%). The transition in conversion between these two extremes occurred at varying arrival frequencies for each catalytic material (*i.e.*, different $BE_B$), with the material described with $BE_B$ of 0.95 eV exhibiting non-equilibrium time-averaged kinetics at the lowest arrival frequencies.

Non-equilibrium time-averaged conversion can be attributed to the contribution of photons promoting the desorption of B*. The contribution of photons was determined by calculating the fraction of desorption events of B* to B(g) resulting from photons (**Figure 6B**). At low photon arrival frequencies, the thermocatalytic reaction equilibrates, as required for a reaction network with all steps exhibiting microscopic reversibility. The photon-promoted desorption of B* is unidirectional and it significantly increases the overall B* desorption rate; the adsorption of B(g) to form B* only releases heat and is not affected by applied photons. The importance of the relative rates of photon-promoted desorption to thermal desorption is further understood relative to photon arrival frequency normalized by the thermal desorption rate constant, $k_{desB}$ (**Figure 6C**). Normalizing the arrival frequency by the desorption rate constant, all catalytic materials (0.75 eV $\leq BE_B \leq$ 0.95 eV) exhibit identical conversion. At the new steady state conversion, the change in the surface coverage of B* equals zero, such that all three contributions to the rate for B* formation sum to zero.

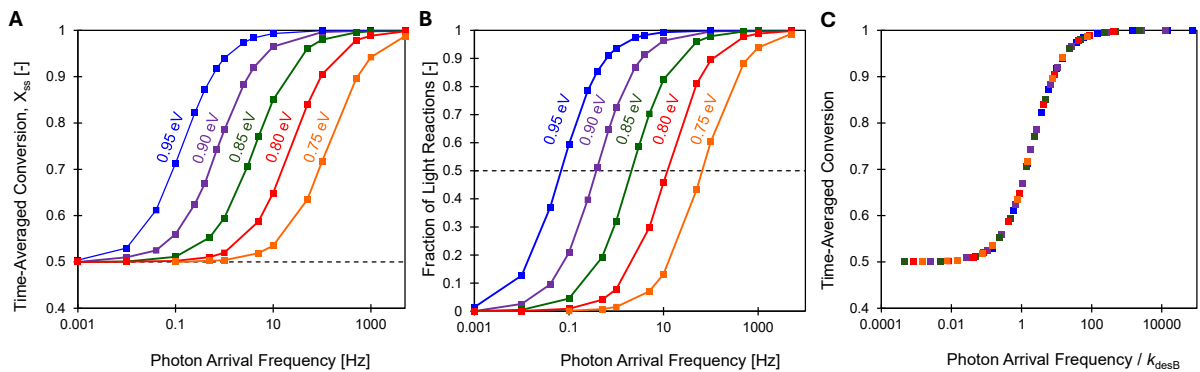

**Figure 6. Non-Equilibrium Conversion of Photon-Promoted Catalysis.** **(A)** Time-averaged conversion of A(g) for binding energy of B* of 0.75 eV (orange), 0.80 eV (red), 0.85 eV (green), 0.90 eV (purple) and 0.95 eV (blue) at 338 K and 100 bar. **(B)** Fraction of reactions to form B* promoted by photons relative to all desorption mechanisms. **(C)** Time-averaged conversion for varying photon arrival frequency normalized the B* desorption rate constant.

$$\frac{d\theta_B}{dt} = r_{A*\leftrightarrow B*} - r_{B(g)\leftrightarrow B*} - r_{B*+ph \rightarrow B(g)} = 0$$

(Eq. 7)

As the photon arrival rate increases, the rate of B* photo-desorption increases, requiring that the rate of surface reaction increase to maintain constant time-averaged conversion. The catalytic surface responds to the depletion of B* through a non-zero net production rate of B*, shifting the new steady-state away from equilibrium

*Temporal Distributions of Photons.* The average time between photons interacting with a catalyst active site is set experimentally by the photon source flux and its relationship to the areal density of catalytic active sites (Eq. 1). However, the distribution of specific times between photons is stochastic and described by a Poisson distribution. As depicted in **Figure 7**, two different photon fluxes with different wavelengths (green and blue) imposed on a catalytic surface (**Fig. 7A(i)** and **7A(ii)**) exhibit comparable photon arrival distributions with different arrival frequencies of 4 Hz and 2 Hz. The lower photon arrival frequency has larger temporal spacing between photons with a distribution tail that extends to longer times (**Figure 7B**); also, the average photon arrival time, τ, is twice as large for the 2 Hz arrival frequency. It is important to note that in both cases, many photons arrive with temporal spacing significantly less than or significantly greater than the average arrival frequency.

The Poisson distributions of a continuous wave laser can be compared directly with the *theoretical* case of a light source with perfectly temporally-spaced photons in **Figure 7A(iii)**. This scenario would prevent the photons from arriving too early before most B* was formed during reaction, and it would also limit the amount of time that B* existed on a surface before a photon arrived. These two benefits should correspond to increased overall rate (*i.e.*, faster conversion of B* as it is formed) and a higher efficiency in the usage of photons arriving at the catalyst surface (*i.e.*, limited number of photons arriving before B* is formed).

To evaluate the impact of controlling the temporal spacing of photons arriving at a catalyst surface, the catalytic reaction of A(g) to B(g) via surface intermediates A* and B* was simulated with a fixed time step (FTS) method. Full details of the simulation are provided in the methods section and supporting information. By this alternative simulation method, photons arrived at the catalyst surface with perfect temporal spacing determined by the preset photon arrival frequency ($\Delta t_{ph} = f_{arrival}^{-1}$). Photons that arrived to a surface and interacted with a site containing B* promoted desorption to B(g), such that the quantum efficiency was 100%. However, photons arriving at the catalyst surface with a site that is open (∗) or contains A* was assumed to be converted to heat, resulting in an *apparent* quantum efficiency, $\eta_{AQE}$, less than unity.

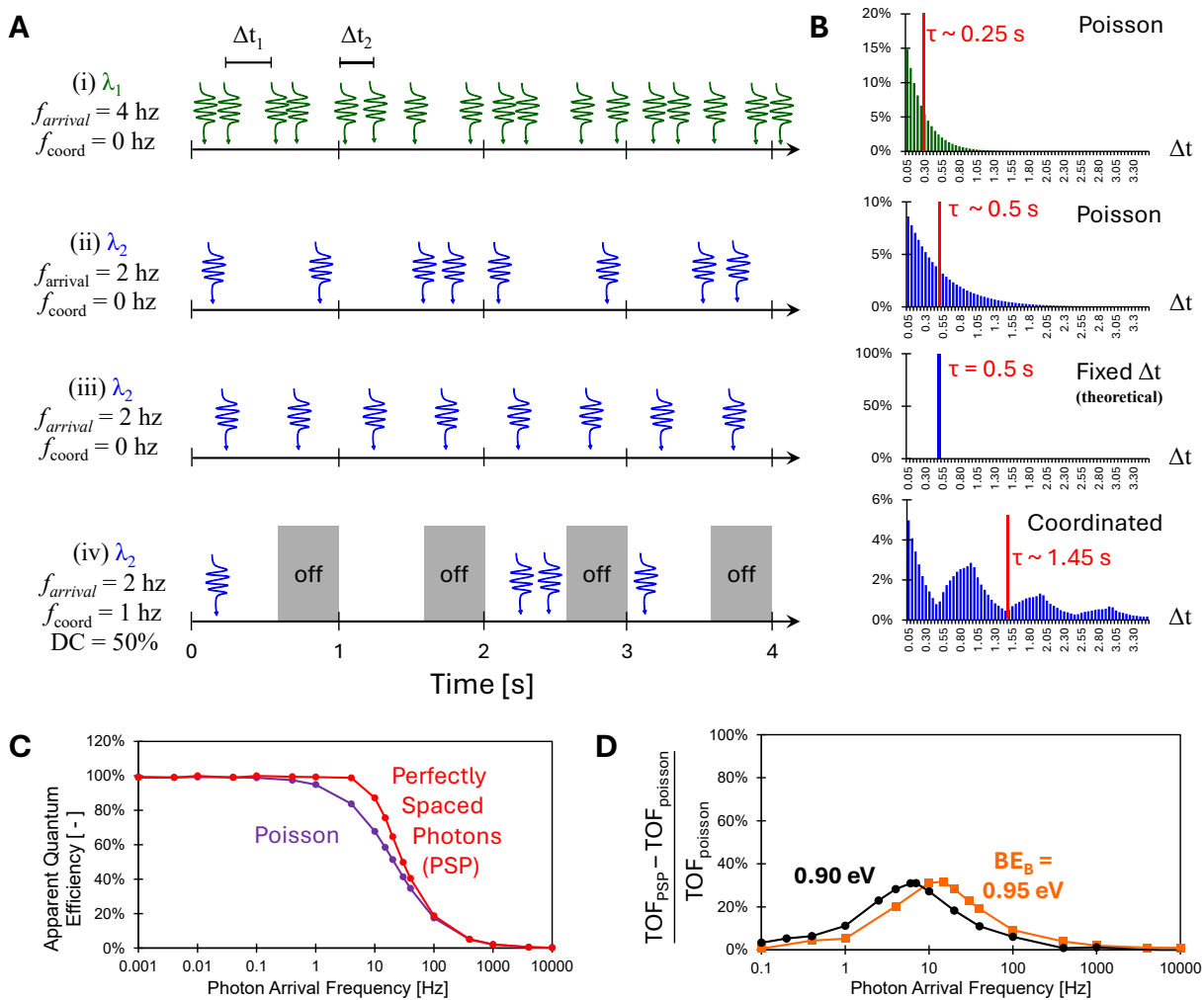

**Figure 7. Photon Arrival Temporal Distributions and Catalytic Efficiency.** **(A)** The illumination of a surface yields a distribution of photons with varying times between arriving photons that varies on average with the average photon arrival frequency, $f_{arrival}$, of 4 Hz and 2 Hz. This is compared with an arrival frequency of perfectly space photons and a fourth system with $f_{arrival}$ of 2 Hz and on/off coordination frequency, $f_{coord}$, of 1 Hz. **(B)** The varying temporal distributions include a Poisson distribution for illuminated surfaces, a theoretical fixed time step, and a more complex distribution resulting from on/off photon source cycles. **(C)** The quantum efficiency of the A-to-B catalytic reaction shifts from high to low efficiency for photon arrival frequencies below and above the reaction rate constant, $k_1 = 9.4$ s$^{-1}$, for 338 K, 100 bar, and 1% conversion for a material at BE$_B$ of 0.95 eV. **(D)** Comparison of the catalytic rate between a Poisson distribution and perfect spacing between photons yields a maximum difference for photon arrival frequencies near the forward rate constant, $k_1$, for a BE$_B$ of 0.90 eV (black) and 0.95 eV (orange).

$$\eta_{AQE} = \frac{Ph[B* \rightarrow B(g)]}{All\ photons} \quad (Eq.\ 8)$$

The apparent quantum efficiency (AQE) was calculated from simulations (**Figure 7C**) applying both perfectly spaced photons and a Poisson distribution for photon arrival frequencies. This particular simulation was conducted at 338 K, 100 bar, BE$_B$ of 0.95 eV, and 1% conversion of A(g) such that the surface reaction forward rate constant was $k_1 = 9.4$ s$^{-1}$. As shown, the two photon distributions exhibit identical apparent quantum efficiencies at photon arrival frequencies far from the surface reaction rate ($f_{arrival} \ll k_1$ & $f_{arrival} \gg k_1$). At a photon arrival frequency of 10 Hz, the AQE of the fixed-time-step photon distribution ($AQE_{FTS}$ = 87.2%) was higher than the AQE of the Poisson distribution ($AQE_{Poisson}$ = 67.7%). Improving the efficient use of photons via control of the their temporal distribution only occurs near the natural frequency of the surface reaction (~$k_1$) and only has moderate benefit (~19% in this case).

Similarly, comparison of the catalytic $TOF_B$ was calculated for simulations (**Figure 7D**) applying both a Poisson distribution and perfectly spaced photons (PSP) for varying photon arrival frequencies and for two different catalytic materials described with a $BE_B$ of 0.90 and 0.95 eV. Plotted as the normalized difference in $TOF_B$ of the two photon distributions (PSP versus Poisson), the maximum difference of only ~30% occurs at the surface forward reaction rate constant, $k_1$. The catalytic $TOF_B$ of the two photon distributions were nearly identical at photon arrival frequencies far from the surface reaction rate ($f_{arrival} \ll k_1$ & $f_{arrival} \gg k_1$).

For both the TOF and the apparent quantum efficiency, the sensitivity to the photon distribution type (PSP or Poisson) is only significant near the surface forward rate constant, $k_1$. At low photon arrival frequencies ($f_{arrival} \ll k_1$), the overall rate of reaction is controlled by the thermocatalytic elementary steps, such that the photon-promoted steps are kinetically irrelevant. Alternatively, at high photon arrival frequencies ($f_{arrival} \gg k_1$), the catalyst surface is inundated with photons, such that the minor differences in photon temporal distribution are kinetically irrelevant; most photons are not utilized for promoting B* desorption. Only for photon arrival frequencies near the reaction rate-determining step ($k_1$) does the subtle differences in the photon temporal distribution impact the catalysis.

*Photon Coordination.* There is limited theoretical benefit for controlling photon distribution, as demonstrated by the light source differences between a Poisson distribution of photons and perfect spacing of photons. Spreading out the arrival times to precise regular temporal spacing between photons only yields small improvements in the apparent quantum efficiency of photons (**Figure 7C**) or the catalytic rate enhancement (**Figure 7D**). Moreover, application of perfectly temporally-spaced photons to active sites is not technically feasible. In this simple A(g) to B(g) system, most of the dynamic benefit (about two orders of magnitude rate enhancement) of the external light perturbation is provided with a simple Poisson distribution of photons (*i.e.*, continuous wave source) promoting a stochastic surface reaction.

The Poisson photon arrival distribution (**Figure 8A**) is a special case yielding an exponential of decreasing probability of temporal spacing between photons arriving to the surface. This distribution calculated via stochastic prediction using random number generation (see supporting information) can be directly compared with the photon-promotion mechanism described using only a reaction rate constant, $k_{ph}$, in a microkinetic model (MKM). As shown in **Figure 8B**, the turnover frequency to produce product B(g) exhibits parity between simulations conducted with either the MKM or kMC method, as calculated at 338 K, 100 bar, and 1% conversion for a material described with $BE_B$ of 0.95. This indicates that in this particular situation, photon-promoted desorption of B* can be described with a rate constant; more complicated photon arrival distributions require more complicated photon distribution models beyond a simple rate constant.

More complicated coordination of light via beam chopping or laser pulsing and its effect on catalytic rate enhancement can be interpreted via the resulting temporal distribution of photons. As shown in **Figure 7A(iv)**, a 50% duty cycle on/off at 1 Hz of a beam of light with photon arrival frequency of 2 Hz yields a complex photon distribution in **Figure 7B**. Rather than a single exponential of a Poisson distribution, the coordinated light distribution exhibits a series of peaks in photon spacing associated with the applied coordination frequency of 1 Hz. The first peak at 1.0 s corresponds to photons arriving at the surface between a single on/off cycle, while subsequent peaks decreasing in size are associated with photon temporal gaps greater than two or more on/off cycles. The potential benefit of coordinating the light via chopping or pulsing is therefore to spread out the arrival times of photons beyond the continuous source and deliver photons at increased probability with specific times that are multiples of the coordination frequency; this is an additional mechanism to control the delivery of light and modulate the extent of chemical perturbation with light.

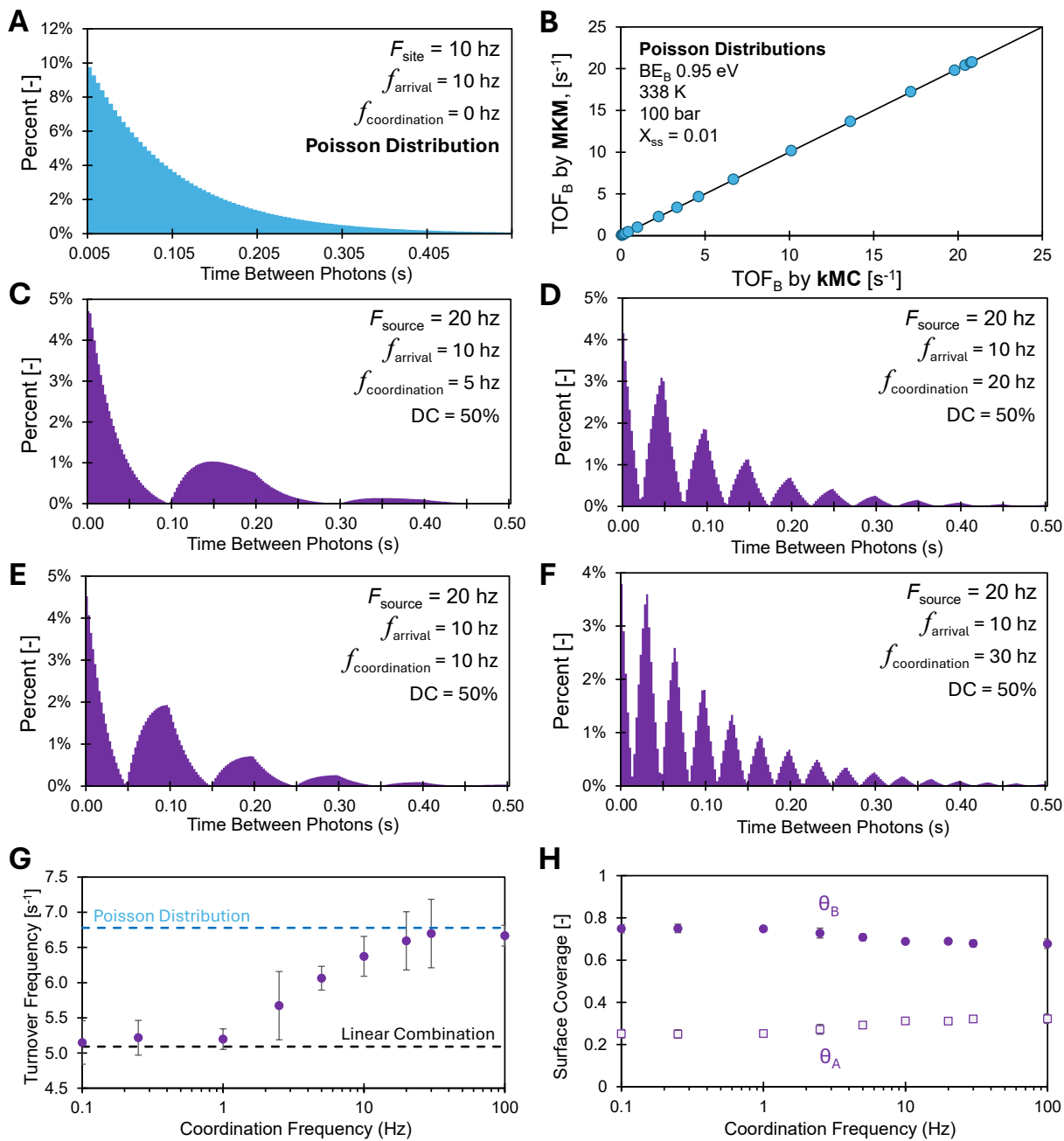

**Figure 8. Coordinated Light Sources for Dynamic Catalytic Rate Enhancement.** **(A)** The distribution of temporal spacing between photon arrival times of an average photon arrival frequency of 10 Hz, $f_{arrival}$, for a Poisson distribution exhibits an exponential decrease in probability. **(B)** The $TOF_B$ simulated at 338 K, 100 bar, and Xss of 1% for varying photon arrival frequencies ($10^{-3} < f_{arrival} < 10^4$ Hz) exhibited parity by two different simulation methods of microkinetic modeling (MKM) and a fixed time step kinetic Monte Carlo (kMC) method with a Poisson distribution of photons. **(C-F)** The photon arrival probability distributions of a light source exhibiting a continuous photon source of 20 Ph site$^{-1}$s$^{-1}$ that is coordinated in on/off 50% duty cycles at 5 Hz **(C)**, 20 Hz **(D)**, 10 Hz **(E)**, and 30 Hz **(F)**. All error bars represent a 99% confidence interval.

The kinetic implications from using coordinated light rather than a Poisson distribution was evaluated via a fixed-time-step kMC simulation that accounted for a light source with on/off duty cycle (0 ≤ DC ≤ 100%) and a coordination frequency, $f_{coord}$, that varied relative to the photon arrival frequency, $f_{arrival}$. Comparisons were made between simulations by fixing the photon arrival frequency, $f_{arrival}$ of 10 Hz, by increasing the source photon flux, $F_{source}$, which can be thought of as site-normalized rate of photon generation from the light source.

$$F_{source} = \frac{f_{arrival}}{DC} [=] s^{-1} \qquad (Eq.\ 9)$$

The photon temporal distributions of four of the considered coordination frequencies of $f_{coord}$ of 5, 10, 20, and 30 Hz are depicted in **Figure 8C-8F** for a duty cycle of 50% and $f_{arrival}$ of 10 Hz. As shown, the number of peaks associated with photon temporal spacing increase as the coordination frequency increases. For low coordination frequencies such as 5 Hz, only two significantly large peaks of photon spacing are observed, while extremely high coordination frequencies such as $f_{coord}$ of 30 Hz exhibit more than 10 significant peaks in photon spacing. A key observation is that as the coordination frequency significantly increases (**Figure 8F**), the photon arrival time probability distribution starts to take on the shape of a Poisson distribution (**Figure 8A**) with an exponentially decreasing probability at long times between photons.

Simulating the catalytic rate of a coordinated beam of light (*i.e.*, chopping or pulsing the light source) was conducted within a fixed time step kMC simulation, as shown in **Figures 8G and 8H**. The simulation calculated the next time for which a photon would arrive at the surface accounting for chopping/pulsing and then simulated up to that point; once the photon arrived at the surface, the additional pathway of photon-promoted desorption of B* was included in the randomly selected possible reaction pathways. The simulations evaluated coordination frequencies over three orders of magnitude (0.1 < $f_{coord}$ < 100 Hz), resulting in three kinetic regimes. At low coordination frequencies ($f_{coord}$ < ~2.5 Hz), the pulsing of light was slow relative to the rate of reaction such that the time-averaged turnover frequency of ~5.1 s$^{-1}$ was equal to a 50% linear combination of the light on condition (10.11 s$^{-1}$) and the dark condition (0.074 s$^{-1}$). This low frequency of source coordination behaved kinetically like two different catalysts, one of which was illuminated and the other which was not. Alternatively, for coordination frequencies above 10 Hz, the time-averaged turnover frequency was equal to the TOF$_B$ resulting from the Poisson distribution of photons from a continuous wave laser. These transitions only resulted in minor changes in the surface coverages of A* and B*; the surface coverage of open sites was negligible under these conditions.

The most unintuitive result of the simulations of coordinated light is the negligible impact of light coordination on the catalytic rate. As shown in **Figure 9A**, a continuous wave laser produces the macroscopic impression of a continuous flow of photons. Alternatively, a light source that pulses or has a beam chopper breaking up the beam would yield groupings of photons, such that macroscopically it would appear that the surface experiences periods of illumination and darkness beneficial for enhancing the catalytic rate. However, the summary of simulation results in **Figure 9B** indicates that coordination of photon sources only has negative impact on the catalytic turnover frequency. For the selected conditions, the steady-state catalytic rate of 0.074 s$^{-1}$ is enhanced beyond the Sabatier volcano maximum almost two orders of magnitude (92x) up to ~6.8 s$^{-1}$ by the continuous illumination of the surface with a Poisson distribution of photons (teal). Shown next are the TOF$_B$ associated with the different considered coordination frequencies (purple), which vary from the rates associated with a linear combination of illuminated/dark rates (~5.1 s$^{-1}$) up to the Poisson distribution rate of 6.8 s$^{-1}$. None of the simulated coordinated light sources exceeded the rate achieved with a continuous wave light source with a Poisson distribution. The limited benefit of coordinating a light source is consistent with prior experiments evaluating the illumination of methanol decomposition on Pt; a continuous stream of photons significantly enhanced the rate of CO(g) formation while chopping of the incoming light stream only provided negligible further rate enhancement.[16]

*Interpretation of Photon Coordination*. The negligible benefit of light coordination via chopping or pulsing is consistent with the inherently dynamic nature of light being comprised of a beam of photons. From the perspective of catalyst active sites on the surface, the arrival of photons occurs at irregular time intervals independent of the macroscopic spacing behavior of photons. For catalytic rate enhancement, the active site requires periods of darkness for adsorption and surface reaction (A* → B*) to occur followed by a period of time where the photon interacts with B* to promoted desorption of B(g). In both of the scenarios of continuous wave illumination and coordinated illumination, these dark/illuminated dynamics are provided. For this reason, stimulation of a catalytic mechanism with light cannot be considered 'static' in the ways that describe other catalyst stimuli such as surface strain or surface charge condensation; surface strain or surface charge condensation can be both static or dynamic while light can only be a dynamic stimulus. In the considered model, the system is the thermocatalytic reaction, and the stimulus is the light source providing photons that periodically alter the thermocatalytic reaction.

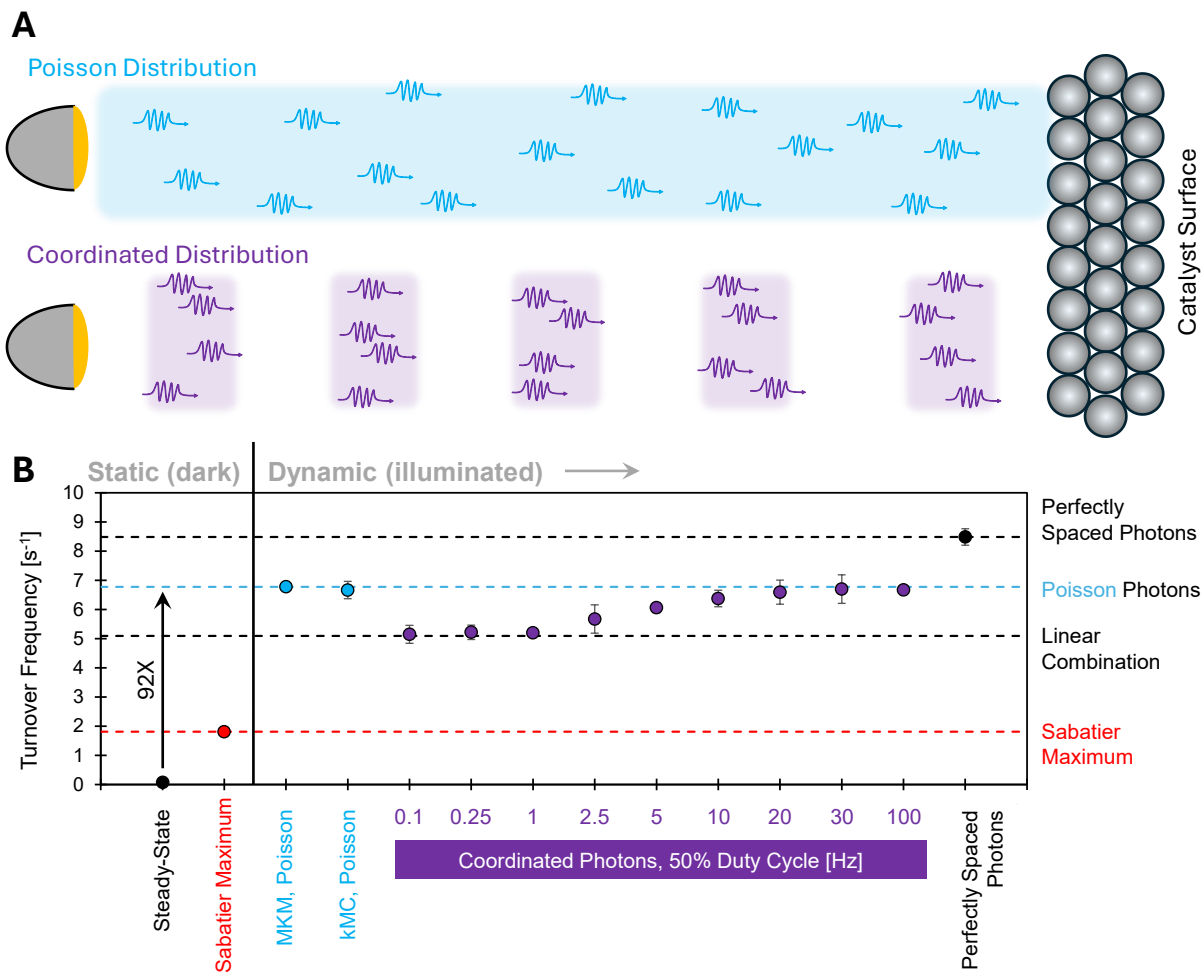

**Figure 9. Summary of Photocatalytic Rate Enhancement.** (A) A light source applying photons to a catalyst surface continuously produces photons as a Poisson distribution, teal, or exhibits coordinated on/off cycles of photons, purple. (B) The time-averaged turnover frequency to form B(g) at 338 K, 100 bar, and Xss of 1% for a material with BE$_B$ of 0.95 eV for several scenarios including: steady state (black), steady state at the Sabatier maximum (red), photocatalysis with a Poisson distribution of photons by two methods (teal), photocatalysis with coordinated photon sources at varying on/off frequencies (purple, 99% confidence interval), and perfectly spaced photons at 10 Hz.

The other consideration of photon chopping or pulsing is the coordination of the surface site states across an entire surface. Photons can only interact locally to promote the desorption of a single adsorbed species over the entire surface.[17] Considering the 'random ensemble' scenario of a Poisson distribution of light imposed on a collection of active catalytic sites, each indistinguishable site experiences discrete events that cyclically transform its state (A*, B*, ∗) with periodicity associated with the surface kinetics and the characteristics of illumination. While all sites traverse through the same of states (adsorbed reactant, transition state, photon-promoted desorption, vacant etc.), each indistinguishable site will be found to exist at any state for any arbitrary moment in time. Across the entire surface, the dynamic rate enhancement provided by photons will exhibit a constant average catalytic rate. In contrast, a 'unified ensemble' scenario would utilize coordinated light such that photons arrived at the surface in unison. Under such a temporally-synchronized environment, the instantaneous rate of the surface changes with time as high rate exists when B* desorbs with the arrival of photons. Low instantaneous catalytic rate exists during the period between waves of arriving photons. The collective instantaneous rate of the unified ensemble is indistinguishable from the instantaneous rate of a single active site. Independent of the level of synchronization between active sites in either the 'random ensemble' or 'unified ensemble,' the overall time-averaged catalytic rate of a surface will be the same, since each active site experiences identical chemical and light-promoted kinetics. In other words, the macroscopic perspective of coordinated light versus a continuous wave illumination of a surface does not identify the dynamic nature of light-promoted catalysis; the dynamic nature of light arises from the temporal characteristics of photon absorption, which occurs at extremely short time scales relative to thermocatalytic reactions.

*Limitations of the simulation.* The conducted simulation was deliberately simplified to an A(g) to B(g) via A* and B* mechanism for the purpose of identifying major kinetic behaviors of the interaction of light and surface chemistry. Comparison of these results requires understanding the limitations of the model, to identify the contributions of real chemical models that could potentially modify global kinetic behaviors. For example, the selected catalytic system exhibits linear scaling of adsorbate and the transition state energies over the entire range of considered binding energies (0.6 eV < $BE_B$ < 1.0 eV); scaling relationships of real reactions are known to deviate from linearity at extreme binding energies.[34,35]

Moreover, the modeled reaction occurs on single sites without a contribution to the binding energies of the total surface coverage.[36,37] Another important assumption in the kinetic model is the assumption that photons only interact with B* and the ratchet efficiency of the photocatalytic reaction is 100%. Defined in equation 10 as the ratio of the rate of B* promoted to B(g) by a photon relative to all photon-promoted reactions,[13,14,38] a high ratchet efficiency indicates that photons can only desorb B* to B(g) without any photon-promoted back reaction to A*.

$$\eta_{ratchet} = \frac{r_{ph:\,B*\rightarrow B(g)}}{r_{ph:\,B*\rightarrow A*} + r_{ph:\,B*\rightarrow B(g)}} \quad (10)$$

Finally, the model does not include transport limitations due to the focus on reaction kinetics, but accelerated surface reaction rates will potentially be limited by diffusion or convection in real particles.[39,40] Despite the simplified reaction model, the general behaviors agree with real photocatalytic reactions that exhibit comparable kinetic rate enhancement in both continuous and chopped catalyst illumination,[16] indicating that the general conclusions of the simple A(g)-to-B(g) photocatalytic model provide insight for real surface chemistries. Future works will build on this initial kinetic model to understand the programmable catalysis of photocatalytic reactions of more realistic surface reaction features.

**Conclusions** The photocatalytic conversion of a generic reaction of A(g) to B(g) via intermediates A* and B* was simulated via the methods of microkinetic models and kinetic Monte Carlo (kMC) for varying temperature, photon arrival frequency, $f_{arrival}$ (*i.e.*, per-site photon flux), and extent of conversion. Simulations evaluating the time-averaged photocatalytic rate for differential conditions ($X_A$ = 1%) demonstrated rate enhancement beyond the Sabatier limit achieving a maximum at the resonance frequency. The resonance frequency ($f_{res}$), defined as the maximum in the effective rate, was shown to equal the forward surface reaction rate constant, $k_1$. Variation of the arrival frequency for materials with binding energies of B* ($BE_B$) greater than the Sabater peak revealed three kinetic regimes: (i) at low $f_{arrival}$, the overall $TOF_B$ was controlled by the rate of thermal desorption of B*; (ii) at high $f_{arrival} > f_{res}$, the overall

TOF$_B$ was controlled by the surface reaction rate defined by $k_1$; (iii) the intermediate kinetic regime exhibited TOF$_B$ at parity with $f_{arrival}$ and a high degree of rate control for the photon arrival frequency. The same three kinetic regimes were observed in an Arrhenius plot simulated for varying temperature, with the intermediate kinetic regime defined by a zero or near-zero slope. Additional simulations via kMC evaluated the non-equilibrium extent of conversion for varying photon arrival frequencies of materials described with different BE$_B$. Deviation from equilibrium was observed for all considered catalytic materials when the photon arrival frequency was comparable to the rate constant, $k_{desB}$, describing the thermocatalytic desorption of B*. The rate of photocatalytic reaction and the apparent quantum efficiency were also evaluated for both a Poisson and a perfectly spaced distribution of photons; the perfectly spaced photons provided an additional minor improvement in both rate and efficiency only for photon arrival frequencies near the resonance frequency. Alternatively, a light source that was coordinated via chopping or pulsing at varying frequency provided negligible benefit in catalytic turnover frequency relative to constant wave illumination of a Poisson distribution of photons. The illumination of the catalytic surface reaction with light indicated significant benefit for enhancing the rate, efficiency and extent of conversion via the dynamic stimulation of surface chemistry with photons.

**Methods.** The general model of A(g) to B(g) via intermediates A* and B* depicted in **Figure 2A** consisting of both thermochemical and photochemical elementary steps was evaluated in Matlab 2019a.[30] It was assumed that thermochemical reactions were unimolecular and reversible through the same transition state, TS, with transition state energy, $E_{a,1}$ or $E_{a,-1}$. Adsorption was defined by the binding energy, BE$_i$, defined as the opposite of the heat of adsorption (BE$_i$ = -$\Delta H_{ads,i}$). The rate constants for forward, $k_1$, and reverse, $k_{-1}$, reactions were calculated with an Arrhenius relationship with transition state theory with pre-exponential constants of $10^{13}$ s$^{-1}$; adsorption pre-exponential constants were $10^6$ (bar·s)$^{-1}$.

The energies of A(g) and B(g) were selected to be zero (0 eV) for an overall reaction that favors neither reactant nor product; the equilibrium conversion, $X_{eq}$, was 50%. The catalyst material of interest (*e.g.*, metal composition) was identified by the descriptor of the binding energy of species B*, BE$_B$. The remaining surface energies were determined by linear scaling parameters. A* was calculated from B* via parameters $\gamma_{B/A}$ and $\delta_{A-B}$,

$$BE_A = \frac{1}{\gamma_{B/A}}[BE_B - (1 - \gamma_{B/A})\delta_{A-B}] \quad (11)$$

where $\gamma_{B/A}$ is the linear slope between B* and A* with offset, $\delta_{A-B}$, defined as the energy where BE$_A$ and BE$_B$ are equal. The transition state energy was calculated with Brønsted-Evans-Polanyi (BEP) linear scaling with parameters α and β using the heat of surface reaction, $\Delta H_{rxn}$, defined as the difference in binding energy of surface species, BE$_A$ - BE$_B$,

$$E_{a,1} = \alpha_1(\Delta H_{rxn}) + \beta_1 \quad (12)$$

Reverse activation barrier, $E_{a,-1}$, was determined by the difference of the heat of surface reaction and the forward activation barrier.

$$k_1 = (10^{13})e^{\left(\frac{-E_{a,1}}{RT}\right)} \quad (13)$$

$$k_{des,A} = (10^{13})e^{\left(\frac{-BE_A}{RT}\right)} \quad [=] \frac{1}{s} \quad (14)$$

The rate of desorption of B* promoted by the arrival of photons was calculated

*Kinetic Monte Carlo Model – Variable Time Step.* The reaction defined in **Figure 6** was simulated kinetic Monte Carlo method via the variable time-step described by Gillespie[41,42,43] By the variable time-step method, each time step was determined by three random numbers ($r_1$, $r_2$, and $r_3$) to determine the duration of time step ($r_1$), the specific elementary reaction ($r_2$), and the quantum efficiency of the utilized photon if selected ($r_3$). The propensity functions, $a(i)$ for each elementary step, $i$, were calculated as the product of the state change vector (υ equal to 1 for all steps) and the elementary first order step rate coefficient, $k_i$ [=] s$^{-1}$. The photon incident frequency per active site, $f_{ph}$, was defined with rate coefficient, $k_{ph}$ [=] s$^{-1}$. The overall propensity, $a_0$, was calculated as the sum of all individual propensities, $a_i$. The variable time step,

$\tau$, was calculated using the total propensity and the first random number,

$$\tau = \frac{1}{a_0} ln\left(\frac{1}{r_1}\right) \quad (15)$$

The selected elementary step, $j$, of the total number of elementary steps, $m$, for each catalyst state was determined as the smallest integer satisfying the following equation,

$$\sum_{j=1}^{m} a_j > r_2 a_0 \quad (16)$$

The third random variable, $r_3$, determined the occurrence of a photon-promoted CO* desorption event, provided this random variable was less than the photon efficiency, $r_3 < \eta$. The simulation was initialized with $t = 0$ and the catalyst in the open site state ($s = 1$) with each time increment, $\tau$, and new catalyst state, $s$, determined by equations 5 and 6. The total number of catalytic turnovers (TON) over the entire simulation was calculated by incrementing +1 for steps 5 and 7 (**Figure 7**) and incrementing -1 for step 6, with the reported simulation turnover frequency (TOF) determined as the ratio of TON by total simulation time. The fraction of light promoted turnovers was determined by counting the total number of occurrences of step 7. The surface coverages were determined as the fraction of time, identified as $\tau$, in each catalyst state. Full model details and algorithm are available in the supporting information.

*Kinetic Monte Carlo – Fixed Time Step.* The 'fixed time-step (FTS) method' imposed a constant duration time step, $\Delta t_{fts}$, calculated as the product of a multiplier, $m_{fts}$, and the inverse of the total propensity, $a_0$, for the kinetic model of **Figure 10**. Additionally, photons were assigned to arrive at the active site separated by constant time increments, $\Delta t_{ph}$. The selected elementary step was determined from the probabilities, $P_i$, of each path with rate coefficient, $k_i$, for the selected step time duration, $\Delta t_{fts}$.

$$P_i = 1 - e^{-k_i t_{fts}} \quad (17)$$

For each time increment, $t + \Delta t_{fts}$, the algorithm assessed whether the elapsed time exceeded the time necessary for the next photon to arrive ($t + \Delta t_{ph}$); arrival of a photon to a catalyst state with B* resulted in instantaneous desorption of B* to B(g). Absent a photon arriving, the selected elementary step, $j$, of the total number of elementary steps, $m$, for each catalyst state was determined as the smallest integer satisfying the following equation,

$$\sum_{j=1}^{m} P_j > r_2 \quad (18)$$

The detailed algorithm and associated Matlab code along with an analysis of fixed step size selection is provided in the supporting information.

*Photon Distribution – Coordinated Light Sources.* The distribution of photons (**Figure 7B**) was calculated by initially calculating a Poisson distribution. By tracking an individual catalytic site, it was then determined if the photon arrived at the site during the on or off period, set by the duty cycle (DC) of the coordinated light source. The detailed algorithm and associated Matlab code is provided in the supporting information.

*Kinetic Monte Carlo – Fixed Time Step with Coordinated Light Sources.* While the fixed time step method initially imposed fixed time step in the simulation and a fixed time between photon arrival, the impact of coordinating the on/off cycles allowed for the fixed time step simulation to vary the time between photon arrival periods. The detailed algorithm and associated Matlab code is provided in the supporting information.

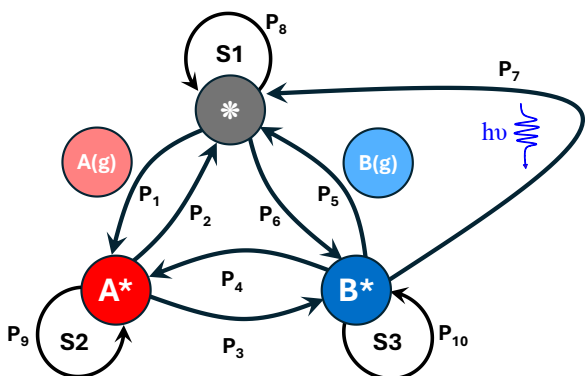

**Figure 10. Thermocatalytic reaction model with photon-promoted desorption.** The reaction of A(g) to B(g) occurs through intermediate surface species of A* and B* through reversible elementary processes. In parallel, photodesorption of B* produces B(g) and an open site, *.

**Acknowledgements.** This work was supported as part of the Center for Programmable Energy

Catalysis, an Energy Frontier Research Center funded by the U.S. Department of Energy, Office of Science, Basic Energy Sciences at the University of Minnesota under award #DE-SC0023464. We acknowledge useful conversations with Professors Matthew Neurock, Omar Abdelrahman, Phils Christopher, Eranda Nikolla, and Michael Gordon.

**Keywords.** Photocatalysis, Resonance, Monte Carlo, Kinetics, Heterogeneous Catalysis

**Supporting Information.** The supporting information is available free of charge.

Reaction model components, reaction rate expressions, Matlab code, figure data sets